\documentclass[12pt,preprint,authoryear]{elsarticle}  

\usepackage{amsmath}    
\usepackage{amsfonts}  
\usepackage{amssymb}
\usepackage{graphicx}

\begin{document}



\title{Comparison of Bayesian Land Surface Temperature algorithm performance with Terra MODIS
observations}

\author{John A. Morgan\\The Aerospace Corporation\\P. O. Box 92957\\Los Angeles, CA 90009}

\begin{abstract}
An approach to land surface temperature (LST) estimation that relies upon Bayesian inference
has been tested against multiband infrared radiometric imagery from the Terra MODIS instrument.
The algorithm employed requires minimal knowledge of surface emissivity, starting from a 
parsimoniously chosen (hence, uninformative) range of prior band emissivity knowledge.
Two estimation methods have been tested.  The first is the iterative contraction mapping
of joint expectation values for LST and surface emissivity described in a previous paper.
In the second method, the Bayesian algorithm is reformulated as a Maximum \emph{A-Posteriori} 
(MAP) search for the maximum joint \emph{a-posteriori} 
probability for LST, given observed sensor aperture radiances and \emph{a-priori}
probabilities for LST and emissivity.

Two MODIS data granules each for daytime and nighttime were used for the comparison. The granules
were chosen to be largely cloud-free, with limited vertical relief in those portions of the
granules for which the sensor zenith angle $| ZA | < 30^{\circ}$. Level 1B radiances
were used to obtain 500 LST estimates per granule for comparison with the Level 2 MODIS LST 
product.  

The Bayesian LST estimators accurately reproduce standard MODIS product
LST values. In particular, the mean discrepancy for the MAP retrievals is
$| \langle \Delta T \rangle | < 0.3 \, K$, and its 
standard deviation does not exceed $1 \, K$.  The $\pm 68 \%$ confidence intervals for 
individual LST estimates associated with assumed uncertainty in surface emissivity are of order 
$0.8 \, K$.

The Appendix presents a proof of convergence of the iterative contraction mapping algorithm.  
The  expectation values of surface temperature in multiple bands, and 
jointly in all bands, converge to a fixed point, within a stipulated convergence criterion.
Provided the support $[T_{min},T_{max}]$ for the calculation of the expectation value brackets 
the maximum in the joint posterior probability for LST, the fixed point coverges to the MAP LST 
estimate in the limit $T_{max}-T_{min} \rightarrow 0$.


\end{abstract}

\maketitle

\section{Introduction}

Land surface temperature (LST) is a vitally important remotely observable tracer of 
mass and energy
exchange across the interface between the atmosphere and the ground.  LST derived from
satellite observations is of interest in its own 
right for local climate studies and climate change monitoring, and as a component of studies of 
land cover, land cover change, surface moisture, and precision farming, among others 
\citep{Dash2002,WanEtAl2004}.
Unfortunately, few surfaces at the bottom of the atmosphere
radiate as blackbodies, and the parameterization of their surface state by means of emissivities
complicates the task of accurate temperature determination.   

This paper continues the development of Bayesian LST estimators that do not require accurate 
knowledge of surface emissivity, 
given radiance in multiple bands \citep{Morgan2005}. The approach may be considered as 
complementary to the widespread use of regression-law based split-window algorithms.

Section \ref{section2} reviews the elements of the Bayesian LST 
formalism.  Two implementations
of the Bayesian approach have been investigated. The first
is the iterative contraction mapping algorithm presented in \citet{Morgan2005}.  The second is
a Maximum-A Posteriori approach to LST estimation.  Section \ref{section3} presents Bayesian
estimates of LST and band emissivity for MODIS imagery.  Section \ref{section4} discusses
results, and a final Section \ref{section5} offers conclusions.  Appendix A presents a proof of 
convergence for the iterative 
algorithm, and estimates of the effect on LST retrieval of uncertainty in aerosol loading.

\section{Elements of Bayesian Land Surface Temperature Estimation} \label{section2}

\subsection{The Posterior Probability for T}

We begin by recalling some of the results obtained in \citet{Morgan2005}, which may be consulted
for details. 
The probability of sensing radiance $I_{i}$ at the top of the atmosphere (TOA), given 
$T,\epsilon_{i},$ and an estimate $\sigma_{i}$ of 
the variance of the sensor radiance noise, is obtained by a 
Maximum Entropy (MAXENT) argument. The MAXENT estimator assumes the existence of a forward model 
for the TOA 
radiance in terms of $T$ and $\epsilon_{i}$.  For the case of a sensor aperture radiance forward model
$I_{i}^{FM}$ that is 
\emph{linear} in band emissivity $\epsilon_{i}$, the posterior probability 
$P(I_{i}\mid T,\epsilon_{i},\sigma_{i})$ is Gaussian in the mismatch between $I_{i}$ and $I_{i}^{FM}$ 
in each band $i$. We also have
\begin{equation}
P(\{I_{i}\} \mid T,\{\sigma_{i}\})=
\prod_{i=1}^{N} P(I_{i}\mid T,\epsilon_{i},\sigma_{i}). \label{eqn60} 
\end{equation}
as the posterior probability for all bands jointly.

Given radiance $I_{i}$ 
in N bands detected at the top of the atmosphere (TOA) that originates from a patch on the 
Earth's surface by a sensor with noise radiance $\sigma_{i}$, the posterior probability that the 
surface is at a temperature T is given by Bayes' theorem as
\begin{equation}
P(T,\epsilon_{i}\mid I_{i},K)=
P(T,\epsilon_{i}\mid K)\frac{P(I_{i}\mid T,\epsilon_{i},K)}{P(I_{i}\mid K)}. \label{eqn3}
\end{equation}
Equation (\ref{eqn3}) is evaluated with aid of 
the prior probability for the surface to be at temperature T and
emissivity $\epsilon_{i}$ given available knowledge $K$,
\begin{equation}
P(T,\epsilon_{i} \mid K) \, dT \, d \epsilon_{i} \propto \frac{dT}{T}\, 
d \epsilon_{i}, \label{eq:prior}
\end{equation}
The result is a posterior probability density.
If one is ignorant of, or unconcerned with, the value of emissivity, it is possible to 
marginalize on the nuisance variable $\epsilon_{i}$:
\begin{equation}
P(T\mid I_{i},K)=\int_{\epsilon_{min}}^{\epsilon_{max}}  \label{nuisance}
d \, \epsilon_{i} P(T,\epsilon_{i} \mid I_{i},K)
\end{equation}
and $P(T\mid I_{i},K)$\footnote{In \citet{Morgan2005},
the posterior probability density was written $P(T\mid I_{i},K) \frac{dT}{T}$} may be
expressed in closed form using error functions. 
The explicit form is given in \citet{Morgan2005}.

\subsection{Land Surface Temperature Algorithms}

We desire to find an estimator for surface temperature using the posterior probability 
given by Bayes' Theorem (\ref{eqn3}).
Two LST algorithms have been developed.  
The first, recounted in \citet{Morgan2005}, 
iteratively computes expectation values of LST and band emissivities from their posterior
probabilities 
while contracting the range of \emph{a-priori} limits on these parameters.  

The second calculational method 
is search for the maximum \emph{A-Posteriori} probability (MAP) value of LST that maximises
$P(T \mid \{I_{i}\},K)$. 
The motivation for investigating a MAP formulation for the Bayesian LST estimator is simple.  The 
original Bayesian LST algorithm presented in \citet{Morgan2005} obtains estimates of LST and band 
emissivity by iterative refinement of 
expectation values for these quantities.  Even though the posterior probability for surface T is
a closed-form expression, this procedure is computationally 
intensive.  Moreover, extensions to the estimator (such as, for example, a more careful treatment of
the forward model, or incorporation of calibration error effects \citep{Morgan2006}) may deprive us
of the comfort of the closed-form solution.  The MAP criterion for an estimator promises to 
dramatically reduce the number of CPU cycles expended per LST estimate, and may yet preserve the
ability to rely upon closed-form solutions.
MAP LST estimates are obtained simply and quickly by a Golden
Section search.\footnote{It is not difficult to extend the MAP approach to
higher dimensions by simultaneously optimizing with Powell's method on (for example)
band emissivity prior limits and LST.  Sample tests of such extensions show promise, but remain to be 
investigated systematically.}

It was conjectured in \citet{Morgan2005} that the iterative algorithm converges in 
general. Appendix A
presents a proof that the iterative algorithm converges, and,
in the limit, converges to the MAP estimate of LST. 
Granted the robustness of the Golden Section search, the existence 
of a solution for the MAP estimator may not appear to be a concern.  However, the MAP algorithm in 
general includes iterative refinement of the \emph{a-priori} limits on surface emissivity.  The 
existence of a unique MAP estimate for LST in that case follows from a straighforward adaptation of 
the method of proof given in the Appendix. 

\subsection{Confidence Intervals}

A desirable feature of an LST estimator is some means of assessing one's confidence in the accuracy 
of individual estimates. 
In the present case LST estimates are obtained by marginalizing over imperfectly known surface 
emissivities. It is naturally of great interest to explore means of judging the effect of an 
uninformative range for the \emph{a-priori} probability for $\epsilon_{i}$.
Direct numerical integration of the posterior probability $P(T \mid \{I_{i}\},K)$ with respect to 
$T$ was used to calculate emissivity prior confidence intervals for LST retrievals presented in this 
study. Suppose that LST uncertainties arising from an uninformative emissivity prior are 
independent of those originating from other sources of error. Then the emissiviy prior confidence 
level may be combined with an estimate of the net effect of all other sources of error by adding in 
quadrature to obtain an overall LST uncertainty estimate.

\section{Application to MODIS imagery} \label{section3}

Both the MAP and iterative versions of the Bayesian algorithm have been applied to MODIS imagery.
This effort took advantage of the availability of Level 1b calibrated radiances and 
Level 2 land surface temperatures (used in lieu of ground truth) for Terra MODIS data granules 
through the Goddard Space Flight Center (GSFC) EOS/DAAC website \citep{Savtchenko2004}.  
Level 2 atmospheric profiles
were obtained from the same site.  The Level 2 surface temperature product is used 
in preference to the standard Level 3 product because it is in the same 1 km
swath format as the Level 1 b radiance data.  
The GSFC-supplied HDFLook display
package was used to manipulate the data granules and associated standard products.

Four MODIS data granules were selected for comparison of Bayesian LST estimates with the 
MOD11\_L2 
LST product.  To the extent possible, 
granules were chosen to have large areas with gentle topographic relief and
minimal cloud cover effects.
A list of the granules used in this study appears in Table 1.
Two granules are daytime, and two nighttime.
The nighttime granule for 2006 day-of-year (DOY) 258 includes the alluvial plains of Iraq and
both the coastal plains and central plateau of Saudi Arabia. 
2006 DOY 268, also nighttime, includes the Great Plains portion of Central Canada.
2006 DOY 350 is a daytime dataset covering Northeast Africa, including large plains in Sudan.
2006 DOY 347, also daytime, covers much of the plains of Central Asia.  The MOD11\_L2 product
reported LST values for approximately $75 \%$ or more of the pixels in all four granules. 
  
Cloud-free pixels for which Level 2 MODIS LST values were reported were selected
for analysis. 
A number of cuts were made in selecting data from the MODIS products for surface 
temperature estimation. 
The use of Level 1b radiance products combined with Level 2 atmospheric profiles to generate LST 
for comparison 
with a Level 2 MODIS temperature product introduces sources of error originating
in, among others, imperfect georeferencing.  Thus one may expect errors in surface altitude to 
affect retrieval 
accuracy.  These effects may be 
exacerbated by use of bilinear interpolation of 4 km horizontally reported profiles to the 1 km 
sampling of the 
spectral
radiances.  In order to minimize this source of potential error, LST retrieval was limited to sensor 
zenith angles
$| ZA | < 30^{\circ}$. Cloudy pixels were masked by rejecting pixels for which the LST 
quality control mask 
bits 0,2,3,4, or 5 were nonzero.  LST retrievals were attempted only for pixels for which the dew 
point temperature
at the base of the water vapor profile exceeded the freezing point of water.

Cuts were also made on the basis of nonphysical or otherwise pathological fields in the granule 
data:  Pixels 
for which the MODIS LST, the MODIS LST error, the surface temperature field in the profile granule 
(otherwise not used in the analysis), the surface air temperature at the lowest level of the profile, 
or the surface pressure were less than zero, were eliminated from processing.  
Finally, pixels for which surface pressure exceeded 1000 millibars, but
for which the surface altitude was less than zero, were omitted from the analysis.

Neither ground truth nor surface meterological range ($VSBY$) is available as a standard product with 
MODIS granules. However, the presence of LWIR bands in the split-window algorithm 
used to produce the MODIS LST product\citep{WanDozier1996,Wan1999} suggests that LST retrievals should
not be too sensitive to the exact value assumed for VSBY in the forward model. Accordingly, this 
study replicates MODIS LST values assuming a nominal, but fairly large, value of 
$23 \, km$ for the surface $VSBY$ parameter for clear pixels. Appendix B 
presents a calculation which supports the contention that, for purposes of the present study,
this approximation should not introduce large errors for clear air pixels.

A random selection of 500 pixels
was made from each granule, subject to data cuts described above.  LST retrievals from both
single-pass MAP and two-pass iterative versions of the algorithm were obtained for all granules.  
In addition, 
$\pm 68 \%$ confidence intervals\footnote{$\pm 68 \%$ confidence levels for lack of emissivity 
knowledge were chosen to approximate the equivalent of  $2 \, \sigma$ intervals, given that 
$P(T \mid \{I_{i}\},K)$ generally displays some skewness. This
should be borne in mind when when comparing with the effect of other error sources described by a 
standard deviation.}
were computed for the MAP retrievals by numerical integration of the posterior probability 
distribution.
The bandset used in the retrievals appears in Table 2.  
The MODerate resolution atmospheric TRANsmission (MODTRAN4) code 
\citep{KniezysEtAl1996,BerkEtAl1999} was used for the forward model calculations, as 
described in \citet{Morgan2005}.  The forward models incorporated MOD07\_L2 air temperature and 
water vapor profiles and assumed a $CO_{2}$ mixing ratio of $382.5$ ppm. 

Initially, the prior probability in (\ref{eq:prior}) was assumed to hold over the range
\begin{equation}
\begin{array}{c}
200 K \le T \le 500 K \\ \nonumber
0.85 \le \epsilon_{i} \le 0.999
\end{array} \label{eqn69}
\end{equation}
for $T$ and $\epsilon$ in each band.  However, it proved advantageous to restrict the range
of the emissivity prior for MODIS bands 31 and 32, 
\begin{equation}
\begin{array}{c} 
0.95 \le \epsilon_{31} \le 0.999 \\ \nonumber
0.95 \le \epsilon_{32} \le 0.999
\end{array} \label{eqn70}
\end{equation}
so as to take advantage of the observation in (Wan and Dozier 1996, Wan 1999) that most natural
land cover types have an emissivity in excess of $0.97$ in these bands. 

In the simulated LST retrievals presented in \citet{Morgan2005}, the equivalent noise radiance 
$\sigma_{i}$ in each band was parameterized by a signal-to-noise ratio (SNR) value obtained by 
approximately inverting the noise-equivalent temperature change ($NE\Delta T$)
for the specified nominal MODIS performance.  Experimentation with actual MODIS
data revealed that the SNR values so chosen greatly underestimated the random variance in 
radiance measured by the actual Terra spacecraft against real terrestrial surfaces, through a real, 
and sometimes dirty, atmosphere.  In consequence, the SNR parameter was reduced from values 
characteristic of
ideal performance, in which the only source of radiance error in sensor measurements is accounted
for by $NE\Delta T$, by a full order of magnitude.  
In fact, the label "SNR" is probably somewhat misleading, since the random 
variability it describes is likely the concatenation of numerous confounding factors, not all of 
which need accurately described as "noise". It is retained because of its use in earlier studies.  

The SNR values used in all retrievals reported in this study appear in Table 2.  These values
were selected on the basis of trial-and-error with pixels from DOY 347 that were drawn 
independently from the selection used for comparison with the MODIS LST product.  
LST estimates are fairly insensitive to the
exact values chosen for band SNR's, so long as these are fairly large.  If the SNR values are too 
large, the noise variance estimate becomes so tight that the retrieval becomes
pathologically overconstrained.  In the most commonly observed pathology, one or more of the 
individual band posterior probability fails to overlap the others; \emph{i. e.} the support of
the joint posterior probability becomes the null set.  If SNR is too small, $\le O(10)$ for MODIS,
LST retrievals become badly biased towards underestimates, and confidence intervals become 
worrisomely large.

Tables 3 and 4 present
the mean, standard deviation and $\chi^2$ of the mismatch for each granule for MAP and 
iterative retrievals, respectively. In Table 3, summary statistics for LST are shown for 
$0.85 \le \epsilon_{31,32} \le 0.999$ as well as for $0.95 \le \epsilon_{31,32} \le 0.999$.
A small number of pixels required relaxation 
of the SNR parameter or emissivity priors in order to obtain an iterative LST estimate,
in a pattern similar to that reported in \citet{Morgan2005}.   

Figures 1-8 display plots of the mismatch between MAP and MODIS LST values, with $\pm 68 \%$ 
confidence intervals, and of mismatch histograms,
for 500 MODIS pixels from each of the four data granules. Overplotted on each mismatch 
histogram is the histogram for the equivalent Gaussian error distribution
\begin{equation}
P(\delta T)=
\frac{1}{\sqrt{2 \pi} \sigma} exp \left 
(-\frac{(\delta T-\langle \delta T \rangle)^2}{2 \sigma^2} \right ).
\end{equation}
Note that the confidence intervals in the plots (whose mean and standard deviation for each granule 
is given in the LST mismatch figure captions) are for the uncertainty in LST resulting from the 
assumed uncertainty prior for surface emissivity only. Thus, they do not (directly) reflect sensor 
noise effects or other sources of uncontrollable variance.\footnote{These other sources of variance 
do enter the confidence interval through the MAXENT estimator for the mismatch between sensor and 
forward model radiance.  Thus, the width of the confidence interval for emissivity uncertainty 
depends upon what value for the SNR parameter is used in the retrieval.} 

\section{Discussion} \label{section4}

Agreement between the Level 2 MODIS LST product and MAP LST estimates obtained in this study
is better than $1 \, K$ in both mean and standard deviation of the mismatch for all granules. 
On average, the mismatch with MODIS LST in Table 3 increases only slightly for the looser 
emissivity prior.  In addition to the tendency for the Bayes LST retrievals to display a slight 
negative bias with respect to the MODIS product (consistent with the observation made in 
\citet{Morgan2005}) the mismatch plots show evidence of a positive trend in the discrepancy with 
respect to MODIS LST.  The trend is readily apparent to the eye for the daytime retrievals, but can 
also be discerned in Figure 3 for (nighttime) DOY 268. MAP and iterative solutions reproduce MODIS 
LST comparably well. Only the DOY 347 retrieval shows a notably different mean mismatch compared to 
MODIS between iterative and MAP approaches. For both algorithms, $\chi^2$ per degree of freedom 
($DOF$) is close to unity.  

The best estimate of MODIS LST accuracy lies in the range 
$\pm 0.6 \,K$ to $\pm 1.0 \,K$ \citep{WanEtAl2004,Guenther2002}. The standard deviation of the
MAP temperature mismatch in Table 3 also lies in this range.  A t-test shows the difference
in means between MODIS and MAP LST values to be statistically significant for both nighttime 
granules and marginally (at the $5 \%$ level) for DOY 347. However, the generally small level of the 
discrepancy indicates that, given better control of error sources, the Bayesian MAP algorithm is 
capable of replicating the accuracy of the MODIS MOD11\_L2 product. 

A small number of retrievals for both algorithms have large discrepancies with respect 
to MOD11\_L2 LST. The cosmetic effect of these retrievals in the mismatch plots may give an 
exaggerated impression of their effect on the accord between MODIS and Bayesian LST.  
Judging from the mismatch histograms, they appear to be outliers resulting from pixels whose 
radiometric accuracy has been compromised 
by one or more of the error sources described earlier, despite imposition of the data cuts.
In the case of the granule for DOY 2006/350, two pixels out of 500 have
a discrepancy in excess of $4 \, K$.  These discrepant values appear significantly out-of-family with 
respect to the largely Gaussian appearance and favorable $\chi^2$ statistic for the retrievals as 
a whole.  In Table 3, the summary statistics for DOY 350 are replicated with the two worst outlier 
pixels removed in the rows marked with an asterisk, with minimal change to the results.

Error sources arising from georeferencing errors in Level 1b radiance products have already been 
discussed.  An additional possible origin for the discrepancies with MODIS LST is errors in the 
forward model. Among other causes, these could result from 
the fixed choice of $VSBY=23 \, km$ (\emph{vide.} Appendix B, however), or occur as a consequence of 
using erroneous values of air 
temperature or water vapor from the MOD07\_L2 profiles in MODTRAN (Wan et al. 2004). Another 
posible source is imperfect cloud 
masking in the Level 2 MODIS products that could corrupt pixel radiances. This last source of 
error would afflict MODIS LST as well, and would have to affect MODIS and Bayesian LST 
differentially in order to contribute to discrepancies.  Finally, the assumed band 
emissivity prior range need not always bracket the actual surface emissivity for a pixel.  

A noteworthy feature of the LST retrievals reported for this study is the comparatively small 
uncertainty for \emph{individual} LST  estimates
originating in the assumed lack of surface emissivity knowledge. The mean value of the
$\pm 68 \%$ interval for individual granules is greatest for DOY 350 (daytime) at 
$0.828 \pm 0.055 \, K$ and least for DOY 268 (nighttime) at $0.638 \pm 0.014 \, K$.
As noted earlier, the emissivity uncertainty confidence interval depends upon the assumed
level of random variance due to noise and other uncontollable factors capable of influencing LST 
retrieval. However, insensitivity of Bayesian LST to emissivity seems likely to hold generally. 
In particular, confidence intervals for  $\epsilon_{31,32} \ge 0.85$ differ little from those for
$\epsilon_{31,32} \ge 0.95$ 

It was remarked earlier that the Bayesian approach to LST estimation was complementary to the more
common split-window algorithmic family.  The MODIS generalized split-window algorithm that
furnished LST values against which performance of the Bayes approach has been tested relies upon
regression against a large training dataset, and requires assignment of pixels to emissivity
classes.  The Bayesian approach, while statistical in its treatment of emissivity uncertainty,
relies upon radiative transfer theory, rather than regression of brightness temperatures against
true LST, to invert radiance to surface temperature.  It is thus a method of remotely determining
LST that (so to speak) minimizes hostages to the fortunes of emissivity knowledge.
The accurate reproduction in the mean of MOD11\_L2 LST in this study suggests the possibility of
using the Bayesian approach for virtual "ground truth" to train local or regional split-window 
LST algorithms.

Although the
golden section search for the MAP version runs faster than the iterative solution in protoype IDL
(Interactive Data Language) code, the difference in execution time is not great, and is in any event 
dominated by the execution
time for the forward model calculations in MODTRAN.  However, the MAP version of the algorithm 
would seem preferable both on the basis of simplicity and because it readily admits extension to 
higher-dimensional searches for optimal joint estimates of LST and band emissivity.

\section{Conclusion} \label{section5}

The performance of the Bayesian land surface temperature retrieval algorithms has been tested in a 
small-scale study by
comparison of Bayesian LST estimates against the MODIS MOD11\_L2 LST product.  
Both the original 
iterative contraction mapping search and a new Maximum \emph{A-Posteriori} algorithm reproduce 
MODIS LST values with good accuracy. The mean mismatch with respect to the MOD11\_L2 LST product for 
Maximum \emph{A-Posteriori} estimates
does not exceed $0.3 \,  K$, and its standard deviation is less than $1 \, K$, for any granule used 
in this study. For the iterative
algorithm, the corresponding mean mismatch does not exceed $0.5 \, K$, nor its standard deviation,
$1.4 \, K$. 

The agreement of the MAP estimates derived from MODIS Level 1B radiance with the MOD11\_L2 LST is 
comparable to the claimed accuracy for MODIS LST. Emissivity prior confidence intervals for 
individual LST estimates shown in Section \ref{section4} show 
that LST retrieval accuracy with the Bayesian approach is insensitive to modest uncertainties in 
surface emissivity.
We conclude the MAP LST algorithm is a viable candidate for LST retrieval in 
circumstances in which band emissivities are poorly known.

It has been shown (\emph{vide.} Appendix A) that the iterative refinement of 
expectation values 
underlying the TES algorithm presented in \citet{Morgan2005} converges, in general. In the limit 
of vanishing support for the LST expectation values, the iterative approach is equivalent to the 
Maximum \emph{A-Posteriori} estimator for LST.  

\noindent{\emph{Acknowledgement:}}

I wish to thank an anonymous referee for \citet{Morgan2005} for suggesting the 
Maximum-\emph{A-Posteriori} approach.

\appendix

\section{Convergence of iterative LST retrieval}  \label{appendixA}
The algorithm used in \citet{Morgan2005} iteratively refines the calculation of the expectation 
values (\ref{eqn61})-(\ref{eqn63}) below
of surface temperature and emissivity in multiple bands by systematic contraction of the
limits of integration. 
It has been found that this procedure, which is described in detail in 
(Morgan 2005), converges rapidly and reliably in practice.   
However, to date there has been no proof that the algorithm actually converges in general.  This
Appendix supplies such a proof.

For reference, the expectation values for T given radiance in bands i,
obtained from
the joint posterior probability for observing radiances $I_{i}, i=1,N$ are
\begin{equation}
\langle T_{i} \rangle = \frac{\int_{T_{min}}^{T_{max}} TP(T \mid I_{i},\sigma_{i})}
{\int_{T_{min}}^{T_{max}} P(T \mid I_{i},\sigma_{i})}, \label{eqn61}
\end{equation}
assuming $T$ is known to lie between a minimum and a maximum,  
while a joint estimator for T given radiances in all N bands is
\begin{equation}
\langle T_{J} \rangle = \frac{\int_{T_{min}}^{T_{max}}TP(T \mid \{I_{i}\},\{\sigma_{i}\})}
{\int_{T_{min}}^{T_{max}}P(T \mid \{I_{i}\},\{\sigma_{i}\})} \label{eqn62}
\end{equation}  
An estimator for the emissivity in band $i$ is given by
\begin{equation}
\langle \epsilon_{i} \rangle =
 \frac{\int_{\epsilon_{min}}^{\epsilon_{max}}\epsilon P(\langle T_{i} \rangle ,\epsilon
 \mid I_{i},\sigma_{i})d\epsilon}
{\int_{\epsilon_{min}}^{\epsilon_{max}}P(\langle T_{i} \rangle ,\epsilon \mid I_{i},\sigma_{i})d\epsilon} 
\label{eqn63}
\end{equation}

\subsection{Assumptions}
Both surface temperature and band emissivities are assumed to lie within limits imposed 
\emph{a-priori}. The surface temperature is limited to the range
\begin{equation}
T_{min} \le T \le T_{max}. \label{eqn:Tlims}
\end{equation}
and emissivity to the range
\begin{equation}
\epsilon_{min} \le \epsilon_{i} \le \epsilon_{max}
\end{equation}
where the minimum and maximum values may be band-dependent.

It is assumed that the maximum of $P(T \mid \{I_{i}\},K)$ is always included in the 
integration. 
In any workable procedure for iteratively refining surface temperature or emissivity by contracting 
the support 
$[ T_{min},T_{max} ]$ of the integrations 
(\ref{eqn61}-\ref{eqn63}) which define $\langle \mbox{\bf{T}} \rangle$ and 
$\langle \epsilon_{i} \rangle$, that support
must bracket the maximum of $P(T \mid \{I_{i}\},K)$ as it contracts. Otherwise, the most 
probable 
values of $\langle \mbox{\bf{T}} \rangle$ and $\langle \epsilon_{i} \rangle$ would be excluded from 
the search, 
in violation of the MAXENT assumption that expected values should maximize the likelihood. 

\subsection{Proof of Convergence}

\noindent \emph{Proposition}: The iterative refinement of (\ref{eqn61})-(\ref{eqn62}) 
converges.
\newline
For clarity we limit the discussion to (\ref{eqn61}) and (\ref{eqn62}). 
The extension to accomodate (\ref{eqn63}) is clear.
A total of $N+1$ expectation values for surface temperature $\langle T_{i} \rangle$ 
(the index $i$ now including the joint estimate) is obtained at each stage of 
iteration.  The values $\langle T_{i} \rangle$ comprise a vector in $\mathbf{R}^{N+1}$.  
These are confined within the subset of
$\mathbf{R}^{N+1}$ delimited by (\ref{eqn:Tlims}) that provides nonvanishing support for the 
integrations in the definitions (\ref{eqn61}) and (\ref{eqn62}).  
Call the vector of expectation values $\langle \mathbf{T} \rangle$.

At the $n$-th stage of iteration, $\langle \mathbf{T} \rangle$ is calculated over the range
$[ T^{n}_{min},T^{n}_{max} ]$ and the mismatch amongst the band LST expectation values 
$\langle T_{i} \rangle$ is scored by
\begin{equation}
\Delta(\langle \mbox{\bf{T}} \rangle) \equiv max |\langle T_{i} \rangle -\langle T_{j} \rangle|
\end{equation}
taken over all pairs of estimates.  
Let the set $\mathbf{X}=\{\mathbf{x}\}$ be the set of all  $m$-tuplets of the form
\begin{equation}
\mbox{\bf{x}} \equiv  \left( \begin{array}{c}  T^{n}_{max}-T^{n}_{min} \\ 
\{ | \langle T_{i} \rangle-\langle T_{j} \rangle |, 
\forall \, i,j \le N+1 \} \end{array} \right)
\end{equation}
consistent with the overall limits (\ref{eqn:Tlims}). Here
\begin{equation}
m \equiv \left( \begin{array}{c} N+1 \\ 2  \end{array} \right) + 1, 
\end{equation}
where the quantity in parentheses is the binomial coefficient.  The set $\mathbf{X}$ is compact, so 
all components of any sequence $\in \mathbf{X}$ are bounded. 
(They are all nonnegative, as well.)  Given points 
$\mathbf{x}^{1},\mathbf{x}^{2} \in \mathbf{X}$ define a metric function
\begin{equation}
d(\mbox{\bf{x}}^{1},\mbox{\bf{x}}^{2}) \equiv \sum_{i=1,N} |x^{1}_{i}-x^{2}_{i}|. \label{eqn:metric} 
\end{equation} 
The pair ($\mathbf{X},d$) can be shown to be
a complete metric space \citep[][p. 35]{AliprantisBurkinshaw1981}.

Define the function $\zeta: \mbox{\bf{X}} \rightarrow \mbox{\bf{R}}$ by
\begin{equation}
\zeta(\mbox{\bf{x}}) \equiv  \theta \sum_{i=1,N} x_{i} \ge 0.
\end{equation}
where $\theta > 1$.

Consider a contractive sequence $x^{n}_{i}$ in $\mathbf{X}$ with $n=1,2,\cdots \,$ for 
which 
\begin{equation}
x^{n}_{i} \ge x^{n+1}_{i}, \, \forall n \label{eqn:contraction}
\end{equation}
Define the mapping 
from one iterate of $\mbox{\bf{x}}$ to another in $\mathbf{X}$ by
\begin{equation}
\mbox{\bf{x}}^{n+1} \equiv M(\mbox{\bf{x}}^{n}). 
\end{equation} 

The function $\zeta: \mbox{\bf{X}} \rightarrow \mbox{\bf{R}}$ is lower semicontinuous,
and $M: \mathbf{X} \rightarrow \mathbf{X}$ is continuous, on $\mbox{\bf{X}}$.
By Caristi's fixed-point theorem \citep{Ok2007} the mapping $M$ has a fixed point 
$M(\mbox{\bf{x}})=\bf{x}$.
The iterative refinement of 
$\langle \mbox{\bf{T}} \rangle$ therefore converges in $\mathbf{X}$.  Moreover,
if 
\begin{eqnarray}
\lim_{n \rightarrow \infty} \, T^{n}_{min} \equiv  T^{\infty}_{min} \nonumber \\
\lim_{n \rightarrow \infty} \, T^{n}_{max} \equiv  T^{\infty}_{max}, 
\end{eqnarray}
then
\begin{equation}
T^{\infty}_{min} \le \langle T_{i} \rangle \le T^{\infty}_{max} \label{eqn:bracket}
\end{equation}  
by the definition of $\langle \mathbf{T} \rangle$. 
$\,\square$

\subsection{Iterative algorithm and maximum \emph{a-Posteriori} probability search}

The fixed point of the iterative mapping drives $\Delta \rightarrow \Delta^{\infty}$,
returning an $N+1$-tuple of LST expectation values, none of which differ from any of the
others by an amount greater than $\Delta^{\infty}$.  It may appear that nothing in the 
result just obtained says any such estimate necessarily gives an accurate value for LST.
In practice, however, it happens that a small number of iterations suffices to yield 
accurate LST estimates.

The manner in which $T_{max}-T_{min}$ contracts remains at our disposal.  We are at liberty 
to force it to tend to zero. We now show that $T_{max}-T_{min} \rightarrow 0$
is equivalent to a contractive mapping chosen with
\begin{equation}
x^{n}_{i} > x^{n+1}_{i}, \, \forall n \label{eqn:recontraction}
\end{equation}
rather than (\ref{eqn:contraction}), and that as $n \rightarrow \infty$
$\langle T_{i} \rangle$ converges to the Maximum \emph{A-Posteriori} estimate 
$T_{MAP}$. 

The support for the computation of $\langle \mbox{\bf{T}} \rangle$ tends to zero as
$x \rightarrow 0$.  There is thus a nested sequence of intervals 
$[ T^{n+1}_{min},T^{n+1}_{max} ] \subset [ T^{n}_{min},T^{n}_{max} ]$.
Cantor's theorem on the intersection of a nested sequence\citep[][p. 32]{AliprantisBurkinshaw1981} 
implies the limit of this process will be a singleton in $\mbox{\bf{X}}$. It was assumed at the 
outset that the support of integrations  $[ T^{n}_{min},T^{n}_{max} ]$ in (\ref{eqn61}-\ref{eqn62}) 
brackets $T_{MAP}$; the singleton thus corresponds to $T=T_{MAP}$, which is also the value of 
$\langle T_{J} \rangle$. We may see this last by the following argument:  In
\begin{equation}
\langle T_{J} \rangle = \frac{\int_{T_{min}}^{T_{max}}TP(T \mid \{I_{i}\},\{\sigma_{i}\})}
{\int_{T_{min}}^{T_{max}}P(T \mid \{I_{i}\},\{\sigma_{i}\})} 
\end{equation}  
as $T_{min},T_{max} \rightarrow T_{MAP}$, we may write
\begin{equation}
\langle T_{J} \rangle = \int_{-\infty}^{\infty} dT \, f(T) T
\end{equation}  
with
\begin{equation}
f(T)= \left\{ \begin{array}{c}
\frac{P(T \mid \{I_{i}\},\{\sigma_{i}\})}{\int_{T_{min}}^{T_{max}}
P(T \mid \{I_{i}\},\{\sigma_{i}\})}, T_{min} \le T \le T_{max} \\
\mbox{else 0}
\end{array}
\right.
\end{equation}
As $T_{max}-T_{min} \rightarrow 0$, it can be shown \citep{Lighthill1960} that 
\begin{equation}
f(T) \rightarrow \delta (T-T_{MAP})
\end{equation}
and thus
\begin{equation}
\langle T_{J} \rangle  \rightarrow T_{MAP}.
\end{equation}

As $x \rightarrow 0$, $\Delta \rightarrow 0$ and the other expectation values 
$\langle T_{i} \rangle \rightarrow T_{MAP}$, a unique converged surface 
temperature estimate.  Conversely, if $T^{n}_{max}-T^{n}_{min} \rightarrow 0$,
$\langle \mathbf{T} \rangle \rightarrow \mbox{\bf{T}}_{MAP}$ by applying the
preceding argument
to each expectation value separately, and $x \rightarrow 0$. Then a 
subsequence $x^{n}$ exists that respects (\ref{eqn:recontraction}). We thus have the

\noindent \emph{Corollary}: In the limit $T^{n}_{max}-T^{n}_{min} \rightarrow 0$,
iterative refinement of unbiased expectation 
values, \emph{i.e.}, those for which $supp  \langle \mbox{\bf{T}} \rangle$ brackets the maximum of 
the $P(T \mid \{I_{i}\},K)$, is equivalent to the 
Maximum \emph{A-Posteriori} (MAP) approach, in which one searches directly for the value of $T$ 
which maximizes the joint posterior probability (\ref{eqn60}).
$\,\square$

This result is likewise
true of the straightforward extension of the proof given here to the case of a contractive iterative
search for a fixed point for $\langle T,\epsilon_{i} \rangle$.  

\section{Insensitivity of LST estimates to meteorological range} \label{appendixB}

The contribution of aerosol loading of the atmosphere to the forward model radiance calculations was 
parameterized by specifying the surface meteorological range, or visibility ($VSBY$).  This is the
variable used by MODTRAN to scale aerosol extinction.
The meteorological range was assumed to be $23 \, km$ for all retrievals presented in
the main body.  Because LST retrievals were only attempted for cloud-free pixels in what were
judged to be reasonably clear air conditions, it seems reasonable to suppose that this assumption
should not introduce significant error so long as few pixels have $VSBY < 23  \, km$.  We now
examine the likely validity of this assumption.  

We begin by adapting the analysis of \citet{Morgan2005,Morgan2006} to obtain
the prior probabilty for surface meteorogical range $VSBY$ 
and use it to construct posterior 
probabilities marginalized over a range of $VSBY$.  The starting point is the apparently trivial
observation that the meteorological range is a \emph{length}:  Traditionally, and loosely, $VSBY$ is
the 
distance at which an observer looking horizontally at a height of two meters just fails to discern 
the contrast presented by an object against the horizon.
In the remote sensing of objects by electromagnetic radiation, Poincar\'{e} invariance requires 
that any two observers must be able to relate their description of events by a 
Lorentz transformation.  In particular, the relation between $VSBY=V$
as observed at time $t$ in primed and unprimed coordinate systems moving with relative
velocity $v=\beta \, c$ with respect to each other along a stipulated line of sight (which might
as well be taken to be parallel to the ground plane)  is given by
\begin{equation}
V=\gamma(V'-\beta \, t') \label{eqn15}
\end{equation}
\begin{equation}
t=\gamma(t'-\beta \: V') \label{eqn16}
\end{equation}
where
\begin{equation}
\gamma=\frac{1}{\sqrt{1-\beta^{2}}}. \label{eqn17}
\end{equation}

The two observers will assign prior probabilities for the occurence of particular values of the
meteorological range 
\begin{equation}
f(V)d \, V \label{eqn10}
\end{equation}
in the unprimed system and 
\begin{equation}
g(V')d \, V' \label{eqn12}
\end{equation}
in the primed one.
In order for observers to agree as to the form of the estimator, the priors must be related by
\begin{equation}
g(V')d \, V'=J^{-1}f(V)d \, V \label{eqn13}
\end{equation}
where
\begin{equation}
J=\frac{\partial{\, V}}{\partial{\, V'}}=\gamma \label{eqn14}
\end{equation}
is the Jacobian for the transformation between descriptions in the parameter space.
Primed and unprimed observers, if equally cogent, must also agree as to the functional form of the
prior probability for $V$, thus
\begin{equation}
f(V)=g(V) \label{eqn42}
\end{equation} 
leading to a form of Schr\"oder's equation for $f$,
\begin{equation}
f(V)=\gamma f(\gamma \: V), \label{eqn43}
\end{equation}
with solution 
\begin{equation}
f(V)=\frac{\mbox{const.}}{V} \label{eqn44}
\end{equation}
for the prior probability.  We conclude that the $VSBY$ parameter, with dimensions of a length,
is a scale parameter, for which the prior probability takes the Jeffreys form (\ref{eqn44})
\citep{Jaynes1968}.

The prior probability (\ref{eqn44}) may be inserted into an estimator for LST which
marginalizes on both surface emissivity and $VSBY$ in each band i,
\begin{equation}
P(T \mid I_{i},\sigma_{i}) = 
\int_{VSBY_{<}}^{VSBY_{>}}
\int_{\epsilon_{min}}^{\epsilon_{max}} P(T \mid I_{i},\epsilon,V,\sigma_{i})d\epsilon 
\, \frac{d \, V}{ V},
\end{equation}
from which a joint posterior probability may be calculated as a function of LST.  This quantity
may be used to compute MAP estimates of LST and confidence intervals.
The estimator will now be used to examine the 
sensitivity of LST estimates when $VSBY$ was allowed to vary over a range of values by means
of an illustrative sample calculation.

LST MAP retrievals and confidence intervals as a function of $VSBY_{<}$ for one of the
pixels used in the DOY 350 granule comparison with MODIS LST 
appear in Figure 9.\footnote{A calculation for a single pixel is given because the use of
nested Romberg quadratures, in which the innermost loop invokes MODTRAN, is  
computationally intensive.}
The $\pm 68 \%$ confidence intervals shown account for the effects of the $VSBY$ and 
emissivity priors only, and include no estimate of uncertainty originating from noise effects.
The upper limit on the visibility prior was fixed at $VSBY_{>}=50 \: km$.  The 
surface T prior for these retrievals was constrained to the range $306 \, K \le T \le 317 \, K$.   

The values of LST and associated confidence intervals for this example 
cannot be directly compared with the MAP value 311.79 K marginalized on surface 
emissivity alone. However, note that the range of LST variation in Figure 9 is limited, even 
within the constrained LST prior limits. Moreover, the confidence intervals tend to a fixed 
range (consistent with the MODIS LST value) as $VSBY$ increases. These observations provide a 
degree of empirical support for the assumption that the LST retrievals presented in the main text 
should not be too sensitive to the exact value of $VSBY$ used to parameterize aerosol loading in 
the forward model, so long as the actual $VSBY$ is great enough.

\setcounter{subsection}{0}
\setcounter{section}{0} 

\,
\,
\,
\,
\,
\,
\,
\,
\noindent{\bf{References}}

\setcounter{section}{0}
\setcounter{subsection}{0}

\begin{table} [p]
Table 1.  MODIS granule datasets 
{\small \begin{tabbing}
.................\=..................................................................  \kill
\rule{80mm}{0.1mm} \\
MOD021KM.A2006258.1900.005.2006260185940.hdf \\
MOD021KM.A2006268.0455.005.2006270054543.hdf \\
MOD021KM.A2006347.0615.005.2006348125833.hdf \\
MOD021KM.A2006350.0830.005.2006352011824.hdf \\
....................................................................................\\
MOD11\_L2.A2006258.1900.004.2006259185248.hdf \\
MOD11\_L2.A2006268.0455.004.2006269162220.hdf \\
MOD11\_L2.A2006347.0615.004.2006349135340.hdf \\
MOD11\_L2.A2006350.0830.004.2006351140447.hdf \\
....................................................................................\\
MOD07\_L2.A2006258.1900.005.2006260192510.hdf \\
MOD07\_L2.A2006268.0455.005.2006270062431.hdf \\
MOD07\_L2.A2006347.0615.005.2006348132045.hdf \\
MOD07\_L2.A2006350.0830.005.2006352015143.hdf \\
\end{tabbing}}
\end{table} 
\begin{table} 
Table 2.  MODIS band definitions, SNR parameters 
{\small \begin{tabbing}
MODIS band...\=.......wavelength/limits\=.................  \kill
MODIS band\>wavelength limits \> SNR \\
\rule{70mm}{0.1mm} \\
    20 \> 3.660-3.840 $\mu m$ \> 25\\
22\> 3.929-3.989 $\mu m$  \> 25\\ 
23\> 4.020-4.080 $\mu m$  \> 25\\
29\> 8.400-8.700 $\mu m$  \> 25\\
31\> 10.870-11.280 $\mu m$ \> 50 \\
32\> 11.770-12.270 $\mu m$ \> 50 \\
\end{tabbing}}
\end{table} 
\begin{table}
Table 3.  Maximum \emph{A-Posteriori} surface temperature results: mean mismatch,
standard deviation, maximum mismatch, and $\chi^2$ per degree of freedom (DOF).  
Results are given for two choices for the
lower limit to the emissivity prior in bands 31 and 32. The rows labeled with an asterisk differs
from those immediately above it by omission of the pixels with the worst two
mismatches compared to MODIS LST
{ \small \begin{tabbing}
Granule.....\=................................\=..............................
\=......................Day............    \kill
Granule \>  $\langle LST  \: mismatch \rangle$  \> $|Max \: mismatch|$ \> $\chi^2 /DOF$ \\
\rule{95mm}{0.1mm} \\
$0.95 \le \epsilon_{31,32} \le 0.999$:\> \>  \> \\
$258.1900$\>$0.238\pm 0.723 \, K$ \>$3.121 \, K$ \> $1.106$ \\
$268.0455$ \>$-0.286\pm 0.416$ \>$3.039$ \>  $1.470$ \\
$347.0615$ \>$-0.058\pm 0.672$ \>$2.790$ \> $1.005$  \\
$350.0830$ \>$-0.036\pm 1.043$ \>$6.974$ \> $0.999$ \\
$350.0830^{*}$ \>$-0.059\pm 0.975$ \>$3.462$ \> $1.002$ \\
..........................................................................................\\
$0.85 \le \epsilon_{31,32} \le 0.999$:\> \>  \> \\
$258.1900$\>$0.454\pm 0.738 \, K$ \>$3.082 \, K$ \> $1.376$ \\
$268.0455$ \>$-0.209\pm 0.429$ \>$2.980$ \>  $1.236$ \\
$347.0615$ \>$ 0.100\pm 0.731$ \>$3.249$ \> $1.017$  \\
$350.0830$ \>$0.216\pm 1.109$ \>$8.866$ \> $1.036$ \\
$350.0830^{*}$ \>$0.189\pm 1.019$ \>$3.737$ \> $1.032$ \\\end{tabbing}}
\end{table} 
\begin{table}
Table 4.  Iterative surface temperature results: mean mismatch, standard deviation, maximum mismatch,
and $\chi^2/DOF$
{ \small \begin{tabbing}
Granule.....\=................................\=..............................
\=......................Day...........    \kill
Granule \> $\langle LST \: mismatch \rangle$ \> $|Max \: mismatch|$ \> $\chi^2 /DOF$ \\
\rule{95mm}{0.1mm} \\
$258.1900$\>$0.263\pm 0.690 \, K$ \>$2.856 \, K$ \> $1.143$ \\
$268.0455$ \>$-0.295\pm 0.425$ \>$3.220$ \>  $1.480$ \\
$347.0615$ \>$-0.670\pm 0.966$ \>$4.081$ \> $1.480$  \\
$350.0830$ \>$0.038\pm 1.355$ \>$7.789$ \> $0.999$ \\
\end{tabbing}}
\end{table} 

\setcounter{section}{0}
\setcounter{subsection}{0}
\setcounter{figure}{0}

\begin{figure} [h]
\includegraphics[width=85mm]{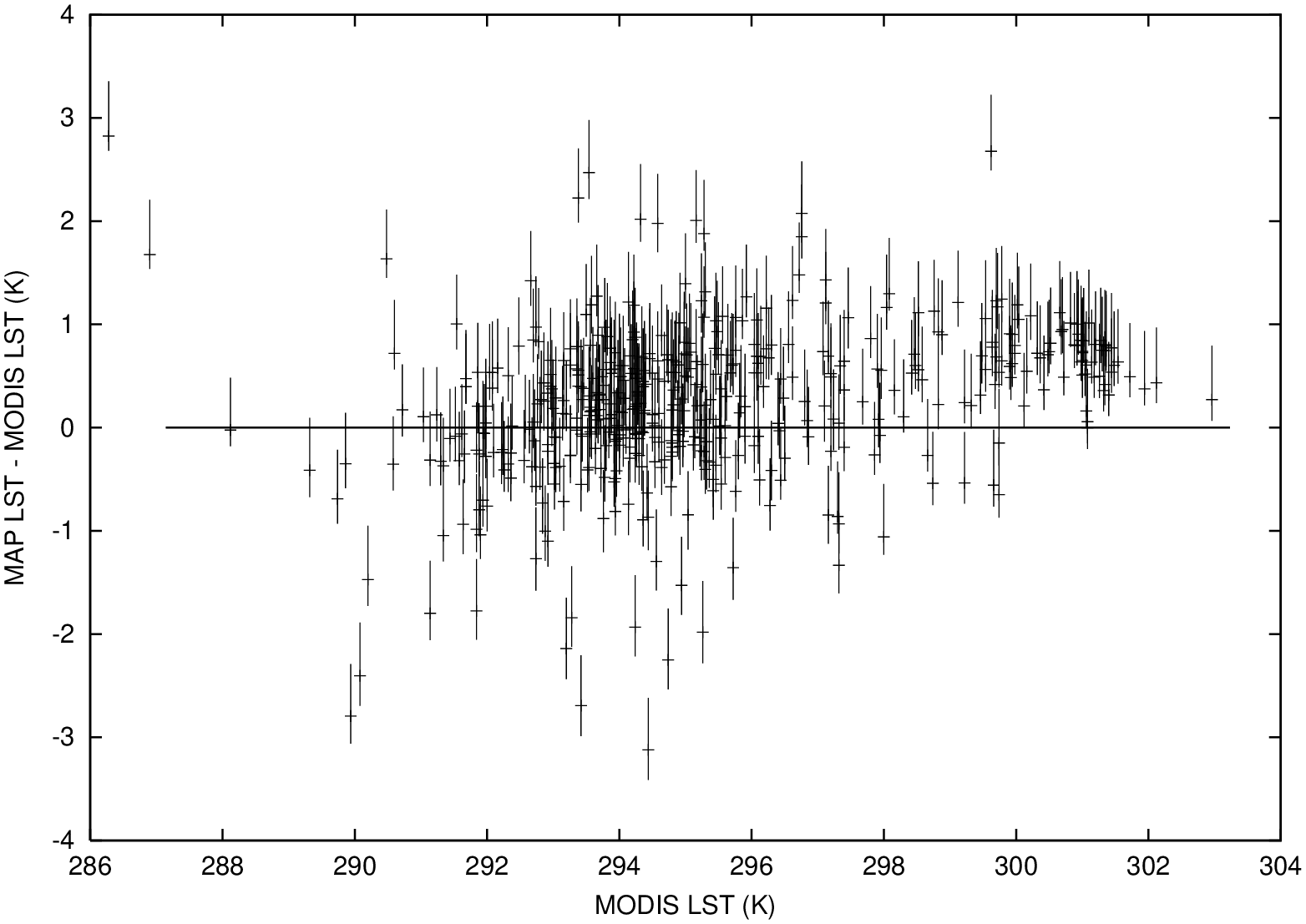}
\caption{Mismatch between MAP and MODIS LST for 500 nighttime pixels 
from DOY 2006/258 granule.  Mean confidence interval 
$\langle \Delta \, T \rangle = 0.730 \pm 0.028 \, K$.
Confidence intervals reflect effects of emissivity uncertainty only.}
\includegraphics[width=85mm]{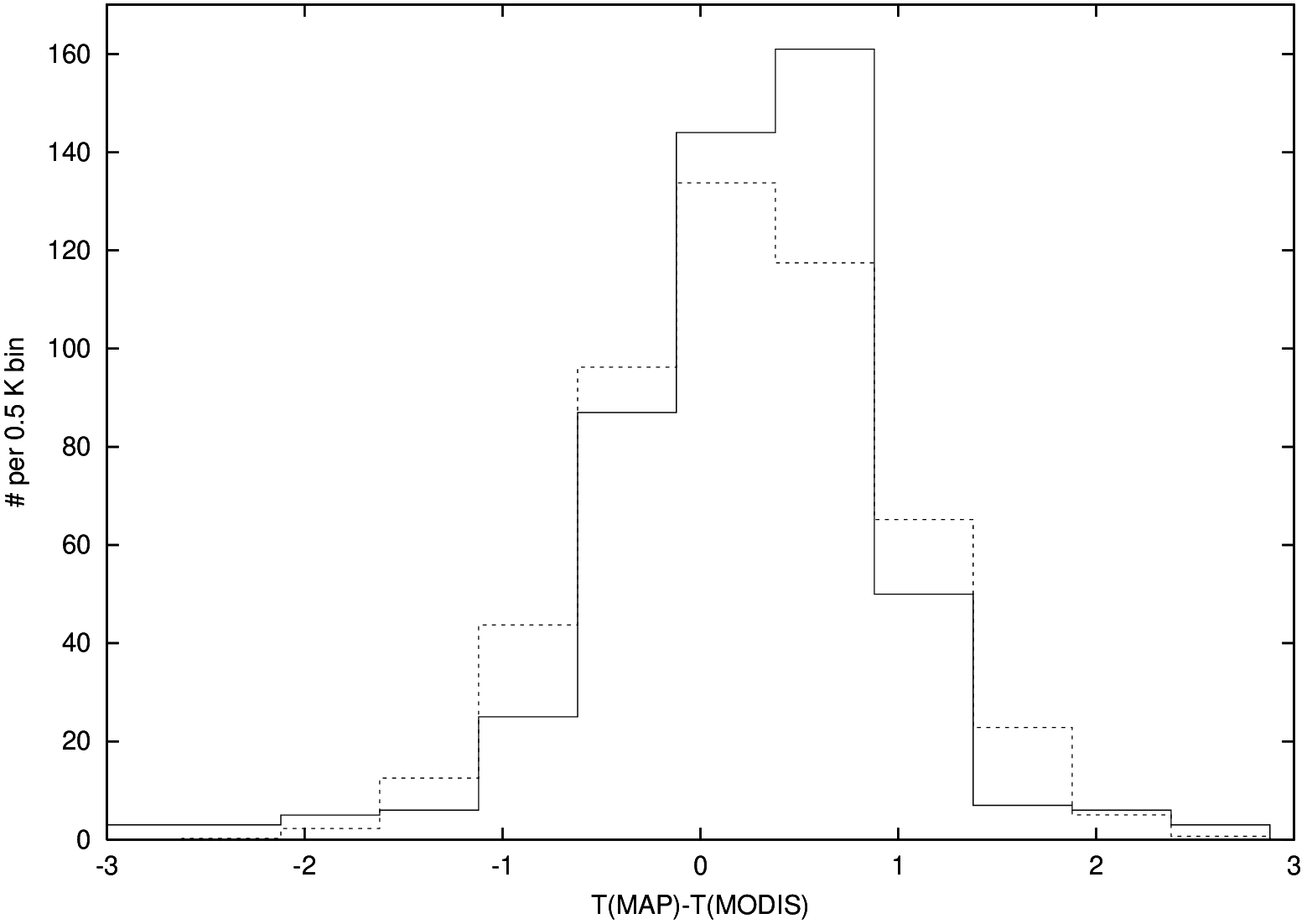}
\caption{Nighttime MAP vs. MODIS LST mismatch histogram for DOY 2006/258.
Solid line: MAP-MODIS mismatch; dashed line: Equivalent Gaussian histogram
corresponding to $\langle \delta T \rangle =0.238 K$, $\sigma = 0.723 K$.}
\end{figure}
\begin{figure} [h]
\includegraphics[width=85mm]{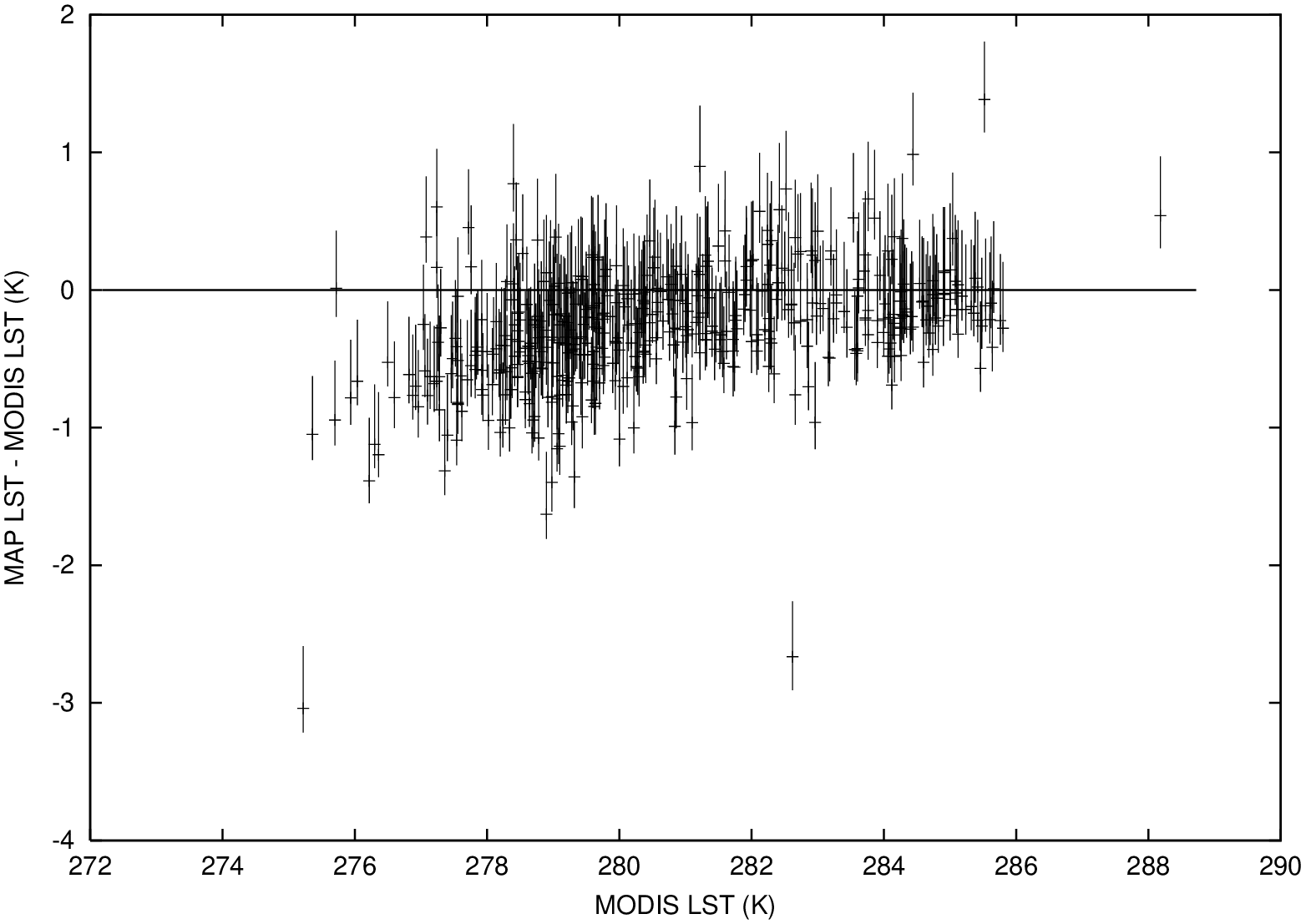}
\caption{Mismatch between MAP and MODIS LST for 500 nighttime pixels
from DOY 2006/268 granule.
Mean confidence interval $\langle \Delta \, T \rangle = 0.638 \pm 0.014 \, K$}
\includegraphics[width=85mm]{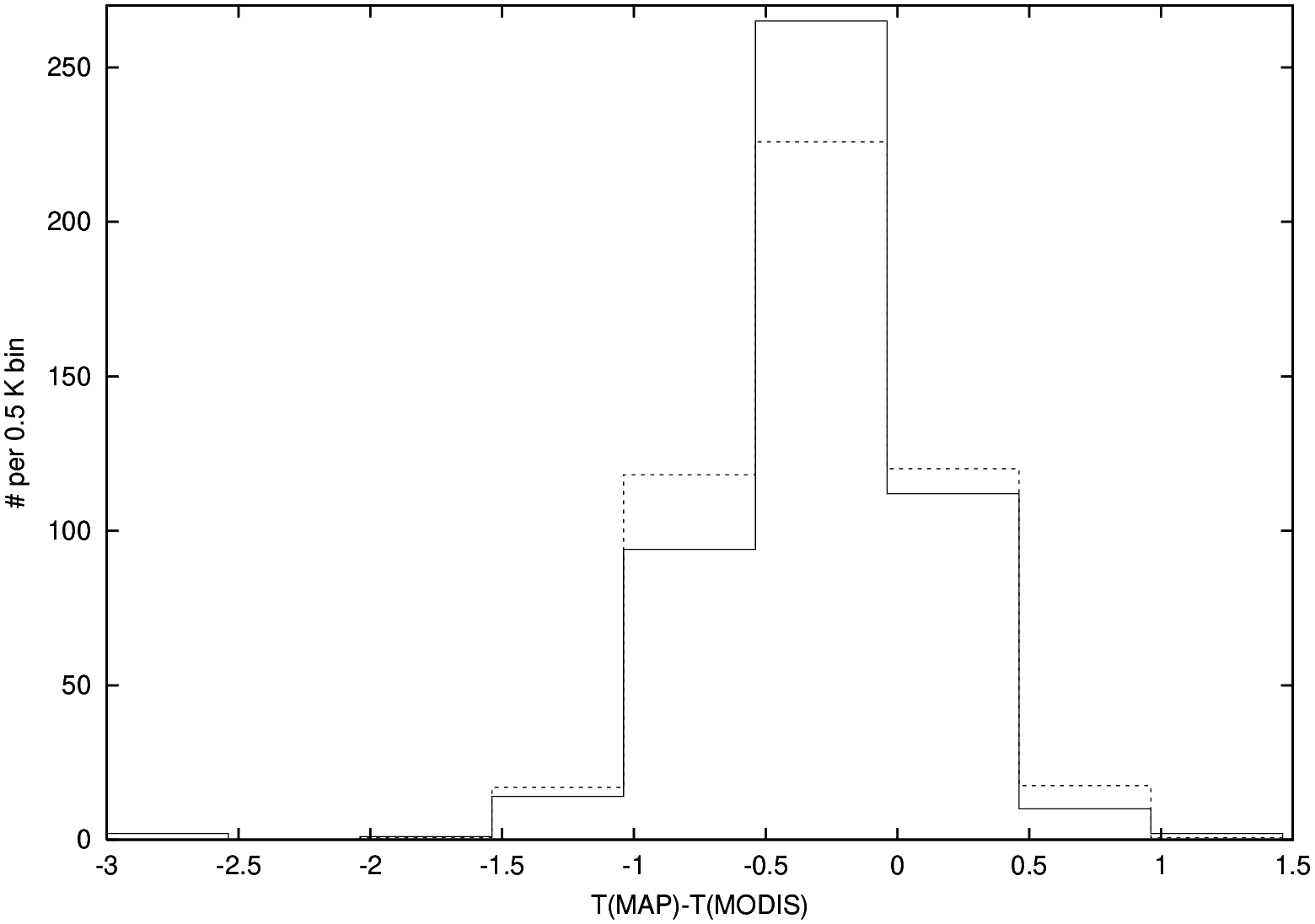}
\caption{Nighttime MAP vs. MODIS LST mismatch histogram for DOY 2006/268.
Equivalent Gaussian histogram: $\langle \delta T \rangle =-0.286 K$, $\sigma = 0.416 K$.}
\end{figure}
\begin{figure}
\includegraphics[width=85mm]{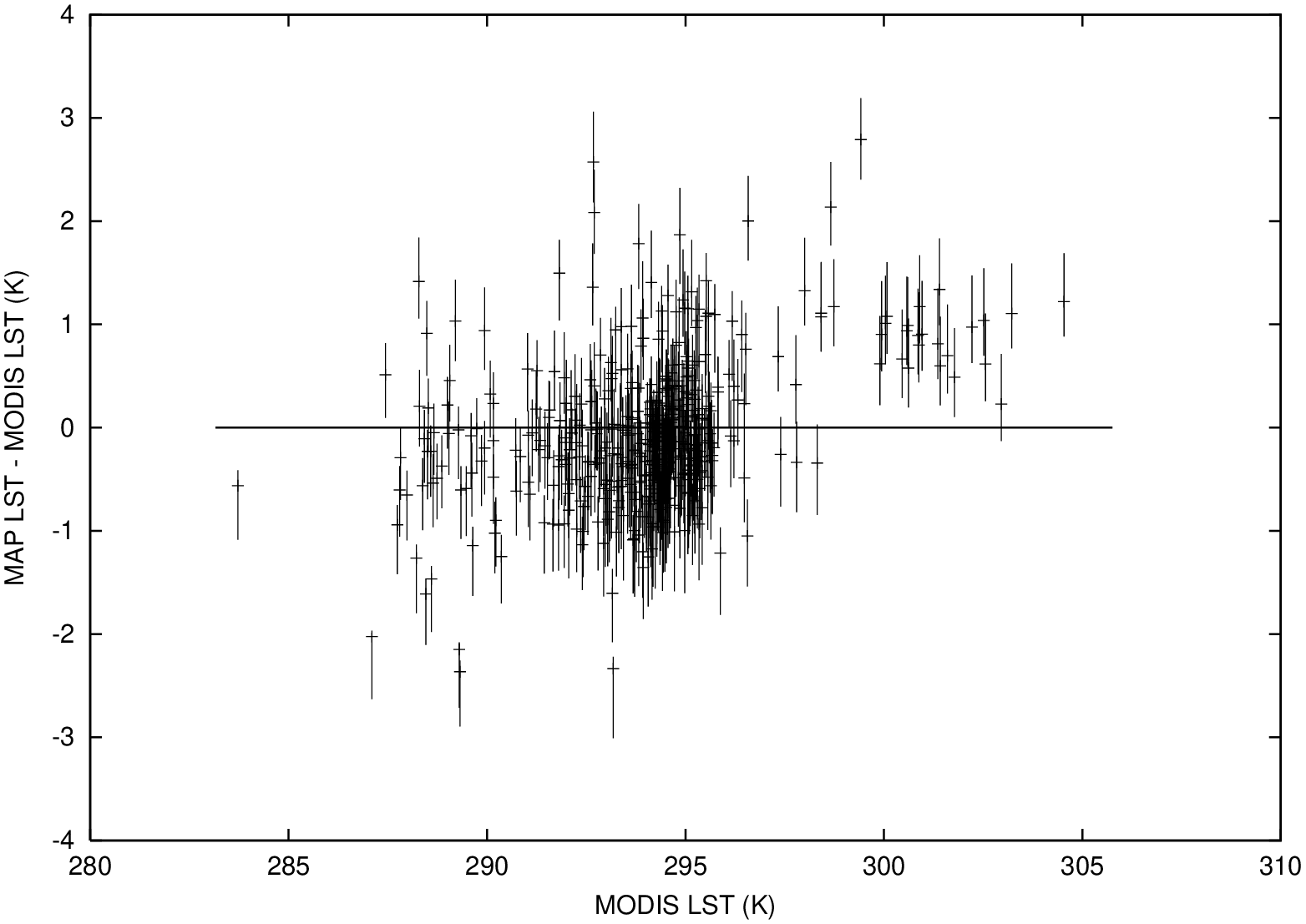}
\caption{Mismatch between MAP and MODIS LST for 500 daytime pixels
from DOY 2006/347 granule.
Mean confidence interval $\langle \Delta \, T \rangle = 0.796 \pm 0.054 \, K$}
\includegraphics[width=85mm]{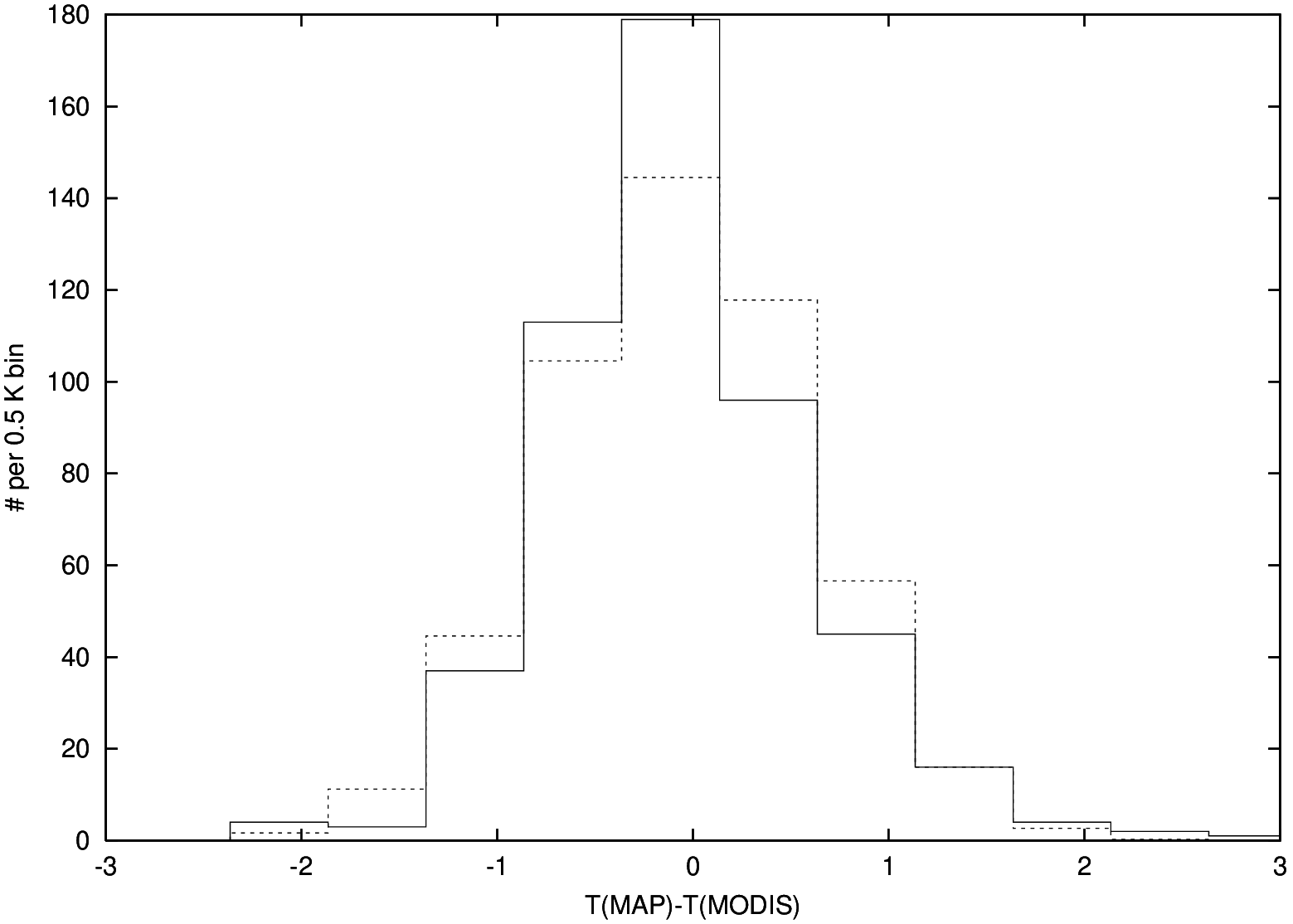}
\caption{Daytime MAP vs. MODIS LST mismatch histogram for DOY 2006/347.
Equivalent Gaussian histogram: $\langle \delta T \rangle =-0.058 K$, $\sigma = 0.672 K$.}
\end{figure}
\begin{figure} [h]
\includegraphics[width=85mm]{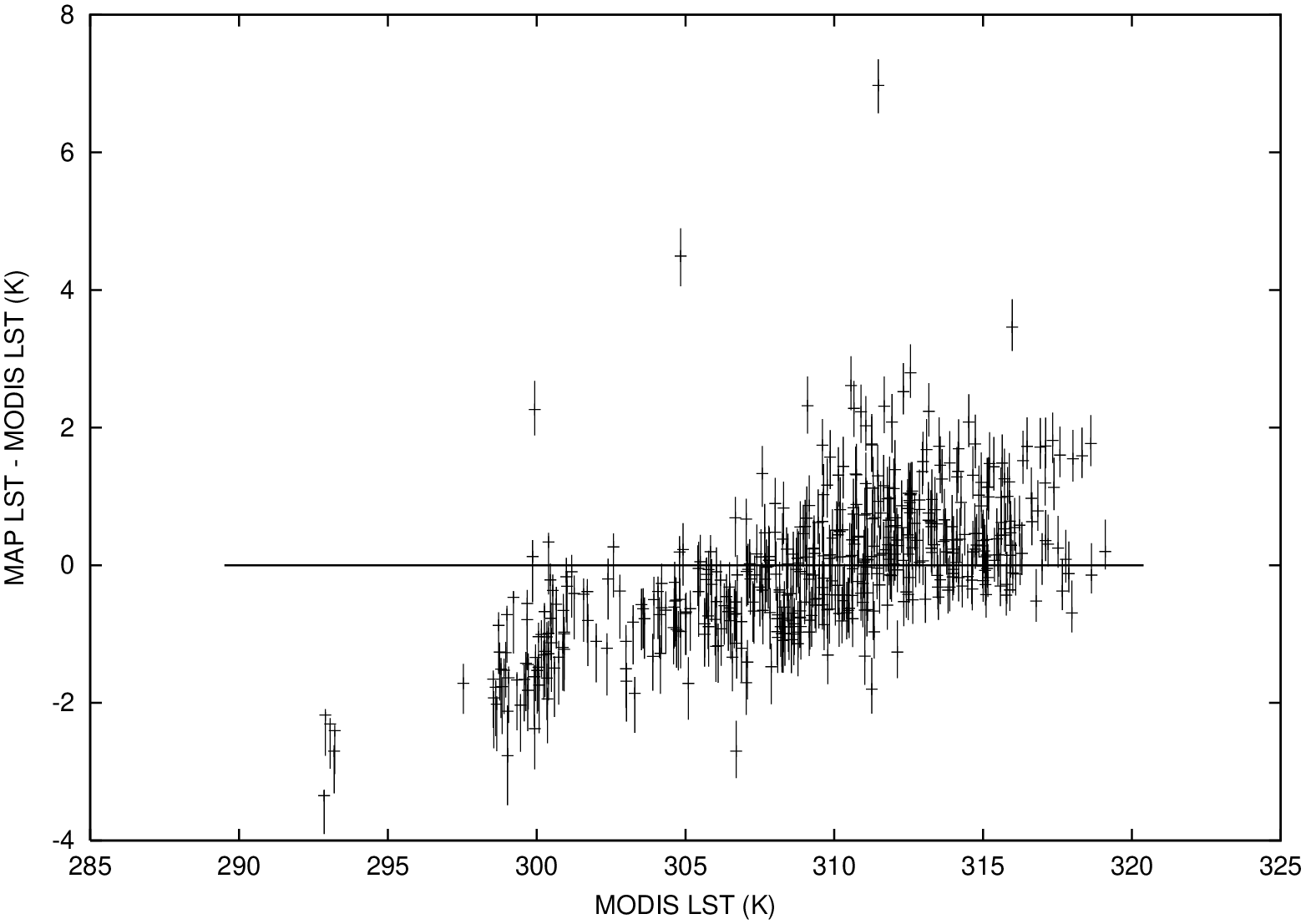} 
\caption{Mismatch between MAP and MODIS LST for 500 daytime pixels
from DOY 2006/350 granule.
Mean confidence interval $\langle \Delta \, T \rangle = 0.828 \pm 0.055 \, K$}
\includegraphics[width=85mm]{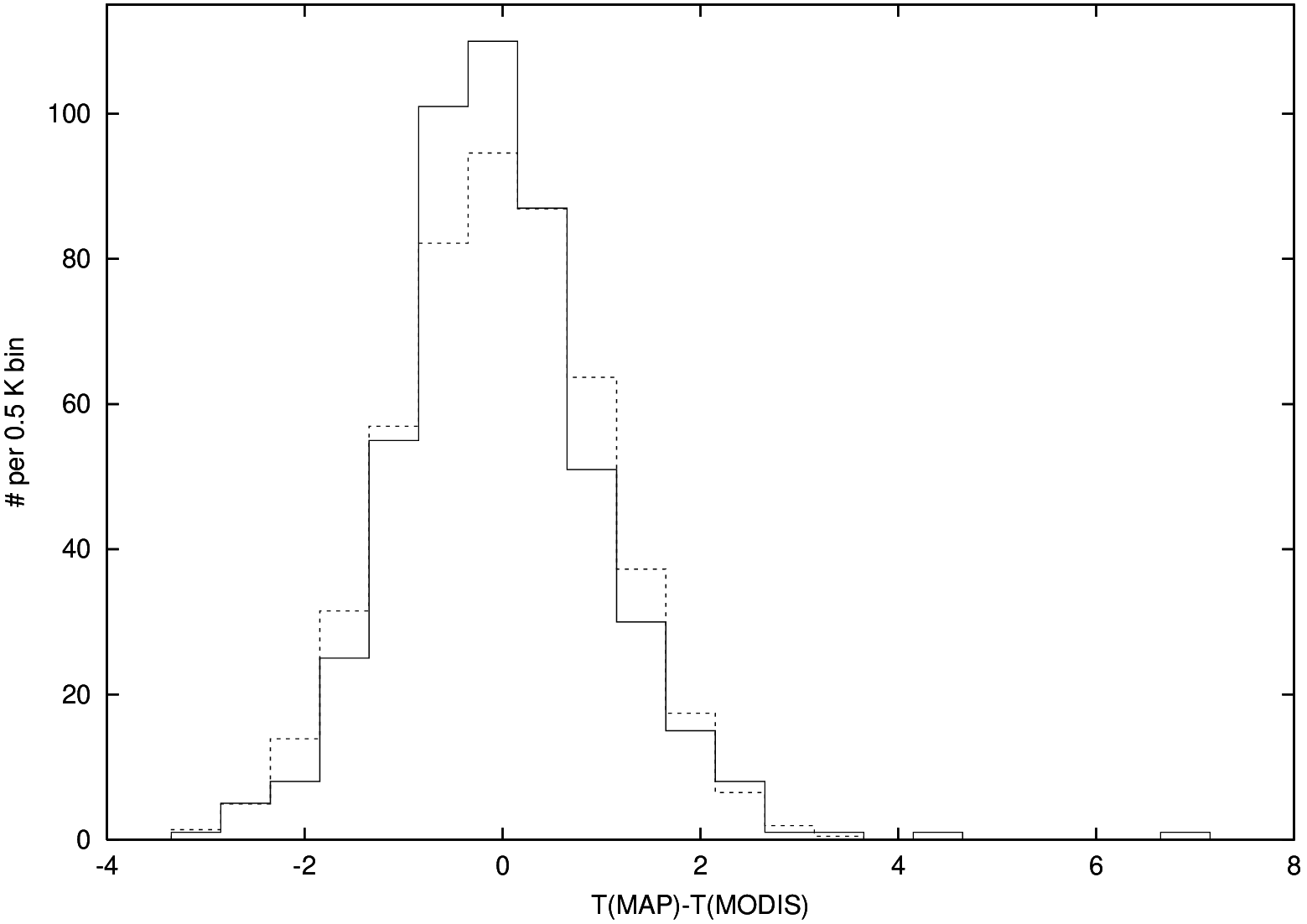}
\caption{Daytime MAP vs. MODIS LST mismatch histogram for DOY 2006/350.
Equivalent Gaussian histogram:
$\langle \delta T \rangle = -0.036 K$, $\sigma =   1.043 K$.}
\end{figure}
\begin{figure} [h]
\includegraphics[width=85mm]{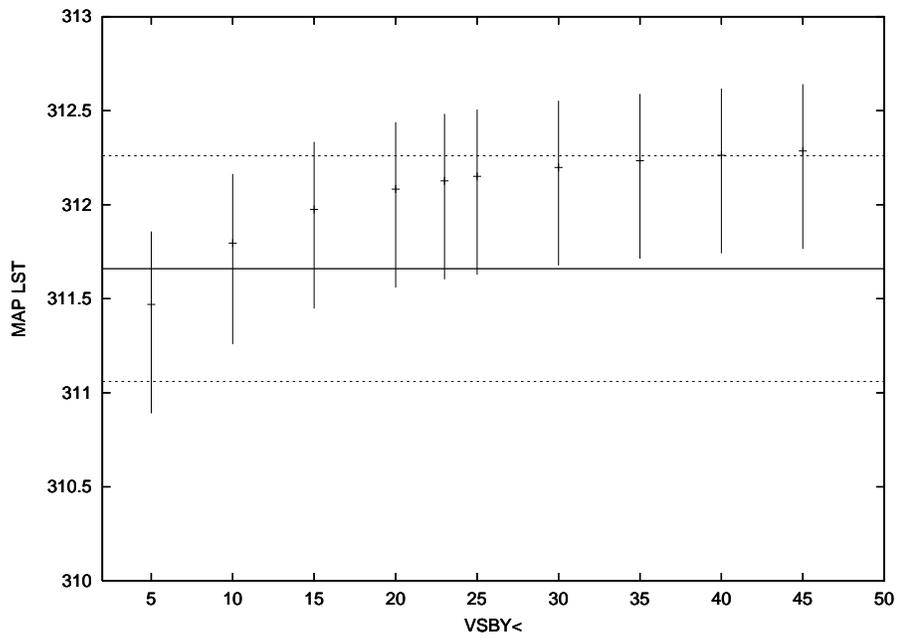} 
\caption{MAP LST estimates, confidence intervals marginalized on emissivity and $VSBY$ 
vs. $VSBY_{<}$ for DOY 2006/350. 
The horizontal lines show the MODIS LST value $311.66 \pm 0.6 K$ for this pixel.}
\end{figure}

\end{document}